\footnotesep \baselineskip	
\everypar={\hangafter=1 \hangindent=3.5em \parindent=0pt \relax}
{ \normalbaselineskip=8pt \parskip 0pt \parindent=0pt
\everypar={\hangafter=1 \hangindent=3.5em \parindent=0pt \relax}
\small
{\bf figure~ #1\/:}
}
{}
\newcommand{\cmsq}{\ifmmode{\rm ~cm^{-2}} \else cm$^{-2}$\fi}
\newcommand{\teff}{\ifmmode T_{\mathrm{eff}} \else $T_{\mathrm{eff}}$ }
\newcommand{\logg}{\ifmmode {\mathrm log}~g \else log\,$g$}

\newcommand{\bmv}{\ifmmode{(B-V)} \else $(B-V)$ \fi}
\newcommand{\ebmv}{\ifmmode{\rm E}_{B-V} \else E$_{B-V}$\fi}
\newcommand{\micron}{\ifmmode\mu{\rm m}\else$\mu{\rm m}$\fi}
\newcommand{\fcgs}{\ifmmode erg~cm^{-2}~s^{-1[B}\else erg~cm$^{-2}$~s$^{-1}$\fi}
\newcommand{\lcgs}{\ifmmode erg~~s^{-1}\else erg~s$^{-1}$\fi}
\newcommand{\flamcgs}{\ifmmode erg~cm^{-2}~s^{-1}~Hz^{-1}\else erg~cm$^{-2}$~s$^{-1}~$\AA$^{-1}$\fi}
\newcommand{\fnucgs}{\ifmmode erg~cm^{-2}~s^{-1}~Hz^{-1}\else erg~cm$^{-2}$~s$^{-1}$~Hz$^{-1}$\fi}
\newcommand{\lnucgs}{\ifmmode erg~s^{-1}~Hz^{-1}\else erg~s$^{-1}$~Hz$^{-1}$\fi}
\newcommand{\kms}{\ifmmode~{\rm km~s}^{-1}\else ~km~s$^{-1}~$\fi}
\newcommand{\mone}{\ifmmode ^{-1}\else$^{-1}$\fi}
\newcommand{\mtwo}{\ifmmode ^{-2}\else$^{-2}$\fi}
\newcommand{\arcsec}{\ifmmode ^{\prime\prime}\else$^{\prime\prime}$\fi}
\newcommand{\arcmin}{\ifmmode ^{\prime}\else$^{\prime}$\fi}
\newcommand{\degs}{\ifmmode ^{\circ}\else$^{\circ}$\fi}
\newcommand{\mv}{\ifmmode {m_{V}}\else${m_{V}}$\fi}
\newcommand{\Mv}{\ifmmode {M_{V}}\else${M_{V}}$\fi}
\newcommand{\msun}{\ifmmode {M_{\odot}}\else${M_{\odot}}$\fi}
\newcommand{\movermsun}{\ifmmode{ \frac{M}{M_{\odot}} }\else{ $\frac{M}{M_{\odot}}$  }\fi}
\newcommand{\rsun}{\ifmmode {R_{\odot}}\else${R_{\odot}}$\fi}
\newcommand{\lsun}{\ifmmode {L_{\odot}}\else${L_{\odot}}$\fi}
\newcommand{\lapprox }{{\lower0.8ex\hbox{$\buildrel <\over\sim$}}}
\newcommand{\gapprox }{{\lower0.8ex\hbox{$\buildrel >\over\sim$}}}

\documentstyle[12pt,emulateapj]{article}  
 
\received{ApJ, Jan 2000}
\accepted{ApJ, 540, Sep 2000}

\lefthead{Green, P. J. et al.}
\righthead{COOL COMPANIONS TO HOT WHITE DWARFS}

\begin{document}

\title{Cool Companions to Hot White Dwarfs}

\vskip0.25cm
\author{Paul J. Green\footnote{Visiting Astronomer, Kitt Peak National
Observatory, National Optical Astronomy Observatories, which is
operated by the Association of Universities for Research in Astronomy,
Inc. (AURA) under cooperative agreement with the National Science
Foundation.} 
}
\affil{Harvard-Smithsonian Center for Astrophysics, 60 Garden St.,
Cambridge, MA 02138} 
\affil{email: {\em pgreen@cfa.harvard.edu}}

\vskip0.25cm
\author{Babar Ali$^1$}
\affil{Infrared Processing and Analysis Center,
Caltech, Mail Code 100-22, 770 S. Wilson Avenue, Pasadena, CA 91125
}
\affil{email: {\em babar@ipac.caltech.edu}}
\and 

\author{R.~Napiwotzki}
\affil{Dr.~Remeis-Sternwarte, Sternwartstr.~7, 96049 Bamberg, Germany}
\affil{email: {\em ai23@sternwarte.uni-erlangen.de}}

\begin{abstract}
Low mass companions to high mass stars are difficult to
detect, which is partly why the binary fraction for high mass stars
is still poorly constrained.  Low mass companions can be detected
more easily however, once high mass stars turn into white dwarfs.
These systems are also interesting as the progenitors of
a variety of intensely-studied interacting binary systems,
like novae, CVs, symbiotics, Ba and CH giants, Feige~24-type systems,
and dwarf carbon stars.

We describe a near-IR photometric search for cool red dwarf companions
to hot white dwarfs (WDs).  IR photometry offers a sensitive test for
low mass main sequence (MS) companions. Our sample of EUV-detected WDs
offers several advantages over previous (largely proper
motion-selected) WD samples: (1) the high WD temperatures
($24000<T_{\rm eff}<70000K$) insure excellent IR flux contrast with
cool dwarfs (2) the range of evolutionary parameter space occupied by
the WDs is considerably narrowed and (3) the random effects of the
intervening ISM provide a complete but reasonably-sized sample.
While some composite systems have been found {\it optically} among
WDs detected in recent EUV All-Sky Surveys, we develop an IR technique
that probes farther down the main sequence, detecting yet more
companions. 

We use detailed DA model atmosphere fits to optical spectra to predict
$K$ magnitudes and distances, against which we contrast our near-IR
observations.  Our photometric survey reveals 10 DAs with a
significant excess in both $J$ and $K$.  Half are newly discovered,
and are most likely previously unrecognized binary systems.  
Neither the frequency of infrared excess nor the mass estimate of the
red dwarf companion correlate with white dwarf mass, as might be
expected if either the EUV detectability or mass of the white dwarfs
were significantly affected by a companion.  Infrared spectra of these
systems should help to determine the mass and spectral type of the
cool companions presumably causing the IR excess, leading to better
estimates of the mass ratio distribution in binaries. 

Counting previously known binaries, and resolved pairs, we find
the total binary fraction of the sample is at least a third. Since
most WD progenitors had initial masses $\geq 2\msun$, we thus provide
a photometric measure of the binary fraction of high mass stars that
would be difficult to perform in high mass main sequence stars.  We
estimate that 90\% of the companions are of type K or later.

\end{abstract}


\section{Introduction}
\label{intro}

EUV-detected stars have revealed in their UV spectra the presence of
about 15 hot white dwarf (WD) companions to bright stars in
non-interacting binary systems  (e.g., Burleigh, Barstow, \& Fleming 
1997). These WDs are hidden at optical wavelengths, because of their 
close proximity to much more luminous companions - main
sequence stars of spectral type K or earlier, or evolved stars. 

Many interacting binary systems where the WD is the primary
(i.e., optically brightest) star have also been found among
EUV-detected systems (e.g., 6 close, interacting white dwarf/red dwarf
binaries by Vennes \& Thorstensen 1994).  Optical or ultraviolet
spectral observations are most commonly used to detect companions to
WD primaries, by searching for (1) the presence of narrow Balmer line
emission overlying the broad smooth Balmer absorption of the WD, (2) a
composite WD + main sequence spectrum, or (3) radial velocity (RV)
variations.  However, only WDs with very close, or intrinsically
active companions will be found by method (1).  For hot WD systems,
composite spectra (2) are only expected to be visible if the
companion's spectral type is early. RV variations (3) require
multiple observations at high spectral resolution, and detection
strongly favors close and/or massive companions.

Such discoveries have been strongly dominated by
these selection effects, with companions biased to earlier types than
predicted by the simulations of deKool \& Ritter (1993) and others.
Scaling from deKool \& Ritter's results, Vennes \& Thorstensen (1994)
estimate that ``at least twice as many close binary systems remain to
be identified from EUV surveys, most of them with a low mass
secondary.'' The resulting sample of binaries known to date therefore
must diverge strongly from the intrinsic distribution, in overall
normalization, as well as in mass and spectral type of the main
sequence companions.  

Large proper motion-selected samples (including those observed in the
IR e.g., Probst 1983a,b and Zuckerman \& Becklin 1992), are
kinematically biased {\em against} luminous (statistically more
distant) young, hot WDs, and span a bewildering range of temperatures,
ages, and metallicities.  Our EUV-selected sample is instead dominated
by freshly-minted disk DAs, providing a narrower slice of parameter
space for stellar evolution (in the wide systems) or for dynamical
evolution (in any new close systems).  Objects known or proposed to
contain white dwarf (WD) stars in currently or previously interacting
binary systems constitute diverse laboratories for the study of
stellar evolution.  A partial list includes novae, cataclysmic
variables, symbiotic stars, Ba and CH giants, Feige~24-type systems
and dwarf carbon stars (Green \& Margon 1994).

We describe here a near-IR photometric survey for low mass companions
to hot white dwarfs (WDs).  By investigating only EUV-detected WDs, we
obtain a very reasonably-sized but complete sample of young WDs, next
to which very late-type dwarf companions can be detected in the
near-infrared by searching for a $K$ excess. Many hot white dwarfs
($T_{\rm eff}>24000$K; Finley et al. 1993) have been detected in the
recent EUV all-sky surveys.  EUV detection of these hot WDs depends
primarily on their temperature, distance, and the intervening Galactic
ISM.  Our sample of EUV WDs (whose selection we define below) offers
excellent flux contrast in the IR relative to optical; cool companions
will almost always be brighter in the $K$ band than the hot WDs.  IR
photometry of hot white dwarfs is nearly non-existent in the
literature.

To know what $K$ mag to expect for the WDs, we benefit from 
constraints on log\,$g$, radius, and $T_{\rm eff}$ derivable
from optical spectra for the WDs in our sample, using 
NLTE model atmosphere fits (Napiwotzki, Green, \& Saffer 1999;
hereafter NGS99).  The resulting predictions for $K$ magnitudes allow
a direct search for any IR excess from a cool companion.  In some
cases, IR colors also provide a preliminary spectral type
for the companion.

\section{Sample Selection Criteria}

We chose to limit our initial sample to DAs, for which model
atmospheres provide
the best temperature and mass constraints.  We start with 73 known DA
white dwarfs in the EUVE bright source list (Malina et al. 1994).  We
exclude sources at low galactic latitudes ($|b|<15$) to avoid crowding
problems, and in the South ($\delta<-20$), yielding a list of 27 DAs.
\footnote{Emission from 40~Eri was resolved with the Einstein HRI, 
and likely arise mostly from 40~Eri~C, the dMe flare star (Cash et
al., 1980), not the white dwarf 40~Eri~B.} 
A similar procedure for DAs listed in the {\it ROSAT} Wide Field
Camera survey Bright Source Catalogue (Pounds et al. 1993) culls an
additional 30 objects.  The Table~\ref{tsample} lists the full sample
of 57, including published data used in this paper.  Model atmosphere
fit parameters for DAs that are not available in NGS99 are taken from
Vennes et al. (1997a) or other references listed in
Table~\ref{tsample}.    

\begin{table*}
\dummytable
\label{tsample}
\end{table*}

Of the 57 sample objects, we present new IR photometry for 47.
We did not obtain new measurements for 10 stars with published
sensitive optical spectrophotometry and IR photometry and/or known
binaries from optical studies
(Feige~24, 
HZ~43, 
GD\,50,
V471~Tau
	(Vennes, Christian, \& Thorstensen 1998),  
PG\,0824$+$289
	(Heber et al. 1993), 
HD\,74389B
	(Liebert, Bergeron, \& Saffer 1990), 
REJ\,1016$-$052, REJ\,1426$+$500 
	(V97), 
and IK~Peg
	(Wonnacott et al. 1993).  
Due to observing constraints (a combination of weather, poor seeing
and faint objects, or celestial placement of objects) no photometry
was obtained for 0138+252 (PG\,0136+251).  
 

On the other hand, we report new IR data for 4 objects that fell
outside the strict sample definitions just outlined.  These include
the known binaries 
REJ\,1036+460 (PG\,1033+464),
REJ\,1629+780 as well as
REJ\,0148$-$253 (GD~1401, also a known binary),  and 
REJ\,0457-280 (MCT\,0455-2812)
which are outside the sample declination limits.  

We note that since the initial sample selection, several relevant
discoveries pertaining to sample objects have been made.
REJ\,0134-160 (GD984) has central Balmer emission components produced by a
dMe companion.  RE\,1440+750 turns out to be a magnetic DA (Dreizler et
al. 1994).  

\section{Observations and Data Reduction}
\label{reduction}

We obtained $J$- and $K$-band photometry of the sample objects at the
Kitt Peak National Observatory\footnote{The Kitt Peak National
Observatory is operated by the Association of Universities for
Research in Astronomy, Inc. (AURA) under cooperative agreement with
the National Science Foundation.} 2.1-m telescope using the InfraRed
IMager (IRIM; Fowler et al. 1988).  The 256$\times$256 IRIM, completed
in 1992, is a HgCdTe NICMOS3 array, which at the $f/15$ focus of the
2.1-m yields a 280$\arcsec\times$280$\arcsec$\ field of view (plate
scale of 1.09\arcsec pixel$^{-1}$).

Table~\ref{tbl-obsrun}~\&~\ref{tbl-obs} summarize our observing
procedure. Table~\ref{tbl-obsrun} lists the individual observing runs
dedicated to this project.  Column~(1) identifies the run.  The
Universal Time (UT) date of the run and the 
photometric standard stars used during each night are shown in
Columns~(2) and (3). Column~(4) identifies
the references containing infrared ({\it J}- and {\it K}- band)
magnitudes for the standard stars, and column~(5) contains a short
overall subjective description for the night conditions.
Table~\ref{tbl-obs} summarizes our observing strategy. Column~(1)
identifies the program object (given as an REJ\,nnnn+mmm).  Column~(2)
lists the nights on which the target was observed, and column~(3) 
any exceptions to our standard observing procedure of 6 dithered
images per target, each image consisting of 6
co-added 10 second exposures.  The exceptions are noted in column~(3) of
Table~\ref{tbl-obs}.  The number of exposures depended on the
faintness of the object.  To minimize any hysteresis effects
in pixel sensitivity, during the observations, we varied the
direction and distance between the position offsets, and never placed
a target at the center of the array.  At least once during each night, 10 - 20
(co-added) dark frames of 10 sec. duration were also obtained.  These
are median-combined to form a single dark frame free of cosmic ray
hits.  The atmospheric seeing (and hence the FWHM of stars) varied
from 1.4--2.5 pixels during the observations.

\begin{table*}
\dummytable
\label{tbl-obsrun}
\end{table*}

\begin{table*}
\dummytable
\label{tbl-obs}
\end{table*}

We reduced our image data using the Image Reduction and Analysis Facility
(IRAF) software.  For each filter separately, the reduction proceeds
as follows.  All images are corrected for any detector non-linearities
(R. Joyce, private communication).  Next, any bad pixels (as defined
by the bad pixel mask; see below) are replaced with the average of the
linearly interpolated values from adjacent pixels in both dimensions
of the image.  Each image is then sky subtracted and divided by the
flat-field.

The sky frame is produced by median-combining all exposures of the
star.  This also serves to subtract the dark current, since the dark
current is included in the sky frames.  The flatfield image is created
as a median-combined frame from a large ($\ge30$) number of individual
images.  Individual images are median-subtracted to account for
sky/background variations.  The flatfield image is dark subtracted and
normalized.  The flatfield image is also used to produce the bad pixel
mask, as defined by the histogram distribution of pixel intensities.
All pixels with intensity values near or beyond the extreme values of
the histogram distribution are defined to be bad.  The bad pixel mask
was produced once and used throughout.

Since our sample is defined to be at high Galactic latitude and hence
uncrowded, simple aperture photometry is adequate.  Following the
arguments presented in Howell (1989) we used a 4.0~pixel diameter
aperture to maximize the signal-to-noise ratio (S/N).  Instrumental
magnitudes are measured for each object using the IRAF task QPHOT with
the radius of the inner sky annulus set to 10~pixels and the width of
sky annulus set to 5~pixels.

For target stars REJ\,0443-034, REJ\,0907+505, REJ\,0940+502, and
REJ\,1442+632, visual inspection failed to reveal the object on
individual images.  In such cases, we shift (by integer pixels) and
combine all the images of the field to form an average mosaiced
image.  Individual frames are median-subtracted prior to averaging to
remove any zero point offsets due to varying sky/background
conditions.  The pixel shifts are calculated from other nearby stars
in the field.  The photometry procedure for these mosaiced images is
described in Appendix~\ref{apx-phot}.

We derive the zero point calibrations for the instrumental magnitudes
from photometric standard stars observed at a variety of airmasses
each night.  These standards stars were mostly selected from the UKIRT
faint standard list (Casali \& Hawarden 1992). The photometric
standards for each night are listed in Table~\ref{tbl-obsrun}.

We eliminated the most extreme outlying measurement for an object if
its elimination reveals it to be $>5\times$ the root-mean-square (RMS)
of the resulting average magnitude.  The resulting relative
photometric error $\sigma$ is taken to be the RMS of all measurements
from each individual frame.  To obtain the final magnitude error, we
add to $\sigma$\ the RMS dispersion in the standard star zeropoints
taken over the night.  The final magnitudes for targets are the mean
values produced by averaging the photometry from all nights on which
the target was observed.  If the target was observed on more than two
nights, we use the algorithm described above to eliminate outlying
measurements.

\begin{figure*}
\plotone{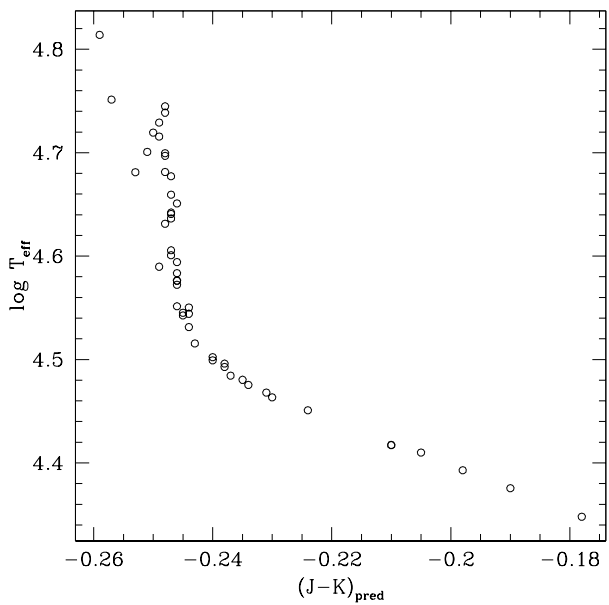}
\caption{Sample DA white dwarf temperature from best-fit model atmospheres
of NGS99 plotted against our predicted $(J-K)$ color for the star
assuming it is solitary.  DA colors are nearly degenerate for
temperatures above about 35,000K.  
\label{f-teffjmk}} 
\end{figure*}

Table~\ref{tirdata} lists for each object with new IR observations
both the observed and predicted magnitudes. Column (1) lists the RE
name, (2) the observed $J$ magnitude and (3) error.  Columns (4) and
(5) list the $K$ magnitude and error, respectively, with the $(J-K)$
color and error in columns (6) and (7).  The absolute magnitude $M_V$
from the best-fit spectral model is shown in column (8).
We list in column (9) the predicted ($V-K$) color based on synthetic
photometry of the best-fit DA models. Predicted near-IR magnitudes for
the DA WDs are listed in columns (10)-(11), and notes in column (12).
We note that our predictions of near-IR {\em colors} for these systems
have very little scatter  over the full sample range, with mean
$(J-K)= -0.24\pm 0.02$, as can be seen from Figure~\ref{f-teffjmk}. 
DA colors - even $(V-K)$ - are nearly degenerate for DA temperatures
higher than about 35,000K.

\begin{table*}
\dummytable
\label{tirdata}
\end{table*}

\section{Analysis}
\label{analysis}

\subsection{Predicting Near Infrared Magnitudes and Colors}
\label{kmags}

We use a grid of 184 NLTE model spectra for DA WDs, with \teff from
20,000 to 100,000 (in 16 steps, coarser toward the the high end) and
\logg\, in 12 equal steps from 7.0 to 9.75.  We predict colors or
magnitudes for each grid model by convolving the model spectrum over
the observed filter bandpasses.  In the visual, we use Mathews \&
Sandage $V$ filter curves and typical CCD quantum efficiency (QE)
curves, while in the near-IR, we use the actual NOAO filter
transmission curves for the $J$ and $K$ filters, along with the IRIM
QE curve. Zeropoints for all magnitudes are calibrated by convolution
with the Vega spectrum of Hayes \& Latham (1975), renormalized (by a
factor of 1.028) to zero magnitude.  As a check on our synthetic
photometry, we tabulated synthetic colors derived from the
spectrophotometric atlas of Bruzual, Persson, Gunn, \& Stryker (BPGS;
Gunn \& Stryker 1983) which includes infrared (IR) distributions from
Strecker et al (1979).  Our synthetic colors agree with published
colors for the corresponding spectral types 
to better than about 0.05mag for blue spectral types (Zombeck 1990).

We predict colors for each program WD star from its
best-fit gravity and temperature using a 2-dimensional linear interpolation
between the colors derived for the nearest grid models.  Predicted colors for
each program star are combined with observed visual DA magnitudes,
to predict their $J$ and $K$ magnitudes, assuming that they
are isolated.  A table of $J_p$ and $K_p$ (predicted $J$ and $K$
magnitudes) for DA stars are listed in Table~\ref{tirdata} along with
the observed magnitudes. 

\subsection{Deriving Distances and Searching for IR Excess}
\label{dist}

Because each of the model spectra are presented in stellar surface
fluxes $F_{\lambda}$ (in cgs units erg~cm$^{-2}$~s$^{-1}$~cm$^{-1}$),
magnitudes we derive for each model in the NLTE grid are effectively
unnormalized surface magnitudes  $V_s=2.5\,{\rm log}\,F_{\lambda}$.
The formula we derive for converting the best fit DA WD surface fluxes
$F_{\lambda}$ (in erg~cm$^{-2}$~s$^{-1}$cm$^{-1}$), surface gravities
log$\,g$, and masses $\movermsun$\, into absolute magnitude is 

$$ M_V = - 2.5\,{\rm log}\,F_{\lambda} 
	 + 2.5\,{\rm log}\,g  
	 - 2.5\,{\rm log}\,\left( \frac{M}{\msun}\right) 
	 + 31.042 $$ 


\noindent 
Our own resulting values of $M_V$ differ little from those for 35 DAs
in common with Vennes et al. 1997a (mean difference $0.01\pm
0.24$mag).  Since $M_V$ correlates strongly with log$g$, this is
consistent with the more detailed comparison performed by NGS99 of
internal, external, and systematic errors in log$g$ and \teff between
the two studies.  For distances to the DA WD, we compare to tabulated $V$
magnitudes, mostly from Table~3 of Marsh et al. (1997).  

We derive the $K$ magnitude of any cool companion, $K_c$, from the
difference between the predicted magnitude of the DA WD, $K_p$, and
the total observed magnitude $K_o$ using

$$ K_c = 
K_p - 2.5\,{\rm log}\left[{\rm dex}\, 0.4(K_p-K_o) - 1\right] $$ 

In Table~\ref{tcompanions} we list the properties of candidate cool
companions detected in our IR-observed sample of DA WDs, estimated
using the combination of the best-fit NLTE models of NGS99, our own
infrared photometry, and published optical magnitudes.  Column (1)
lists REJ name. The companion $J$ and $K$ magnitudes calculated as
detailed above are listed in columns (2) and (3).  Spectral type
of the candidate 
companions can be crudely estimated from the derived $M_{K_c}$ or from
$J-K$.  Since the combined errors on $M_{K_c}$ are in general lower, we
list this in column (4) and the resulting spectral type estimates in
columns (5) and (6), assuming young or old disk population for
the red dwarf.  The implied $V$ magnitude (from $V-K$ for the
old disk spectral type) is shown in column (7), with notes in column 
(8). 

\begin{table*}
\dummytable
\label{tcompanions}
\end{table*}

\subsection{Candidate Cool Companions}
\label{detect}

Our adopted criteria for claiming a significant near-IR excess over
the predicted values for an isolated DA is simply a 3$\sigma$ excess in
both $J$ and $K$ relative to the predicted IR magnitudes of the DA white
dwarf, $J_p$ and $K_p$.  Errors in our observed $K$ magnitudes are
generally 2-3 times higher than for $J$, but provide greater
wavelength contrast to the optical.  We do not use ($J-K$) as a
discriminant because its combined errors are highest.
The errors we adopt for the excesses $(J_p - J_o)$ and $(K_p - K_o)$ 
include our photometric errors in $J$ and $K$, combined with  assumed
errors of 0.1mag each on the predicted values of $K_p$ and $J_p$.  

As an example, for REJ\,1845+682, we calculate from the 
published $V$ magnitude and model atmosphere fits in
Table\,1 a predicted magnitude with its estimated error of
$K_p=16.2\pm0.1$ (Table\,4).  The difference between this and the
observed magnitude $K=14.37\pm0.19$ (Table\,4) is thus
$K_{excess}=1.83\pm0.21$mag, qualifying as an $8\sigma$ excess. 

Of the 47 objects with reliable $K$ photometry
(which excludes REJ\,1235+233/PG1232+238) we find 10 systems with a
significant IR excess. Of these, 5 are known binary systems.  Of the 5
that are newly recognized systems with IR excess, 4 have spatially
resolved objects discernible either in the near-IR or optical
images within 10\arcsec.  In only some cases, noted in
Table~\ref{tcompanions}, these objects are included in the near-IR
photometry. 

The overall properties of DAs with IR excess reveal no obvious bias
within our observed sample.  The mean distance of the DAs without
excess is $144\pm98$pc while those 11 with excess have $117\pm53$pc.
Similarly, the mean values and distributions of \teff, \logg, $V$, and
$M_V$ of the DAs without excess are indistinguishable from those 11
with excess.

\begin{figure*}
\plottwo{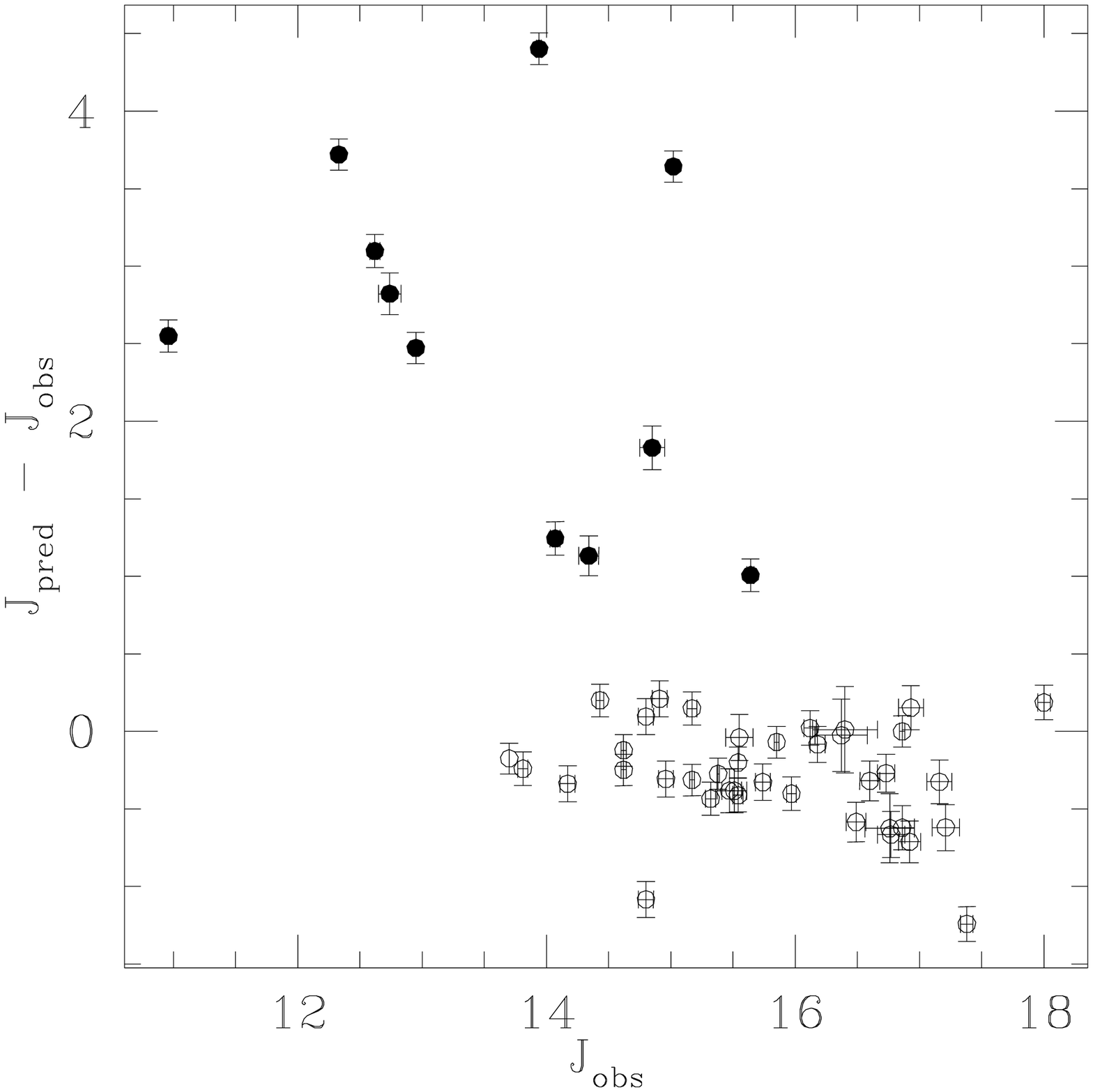}{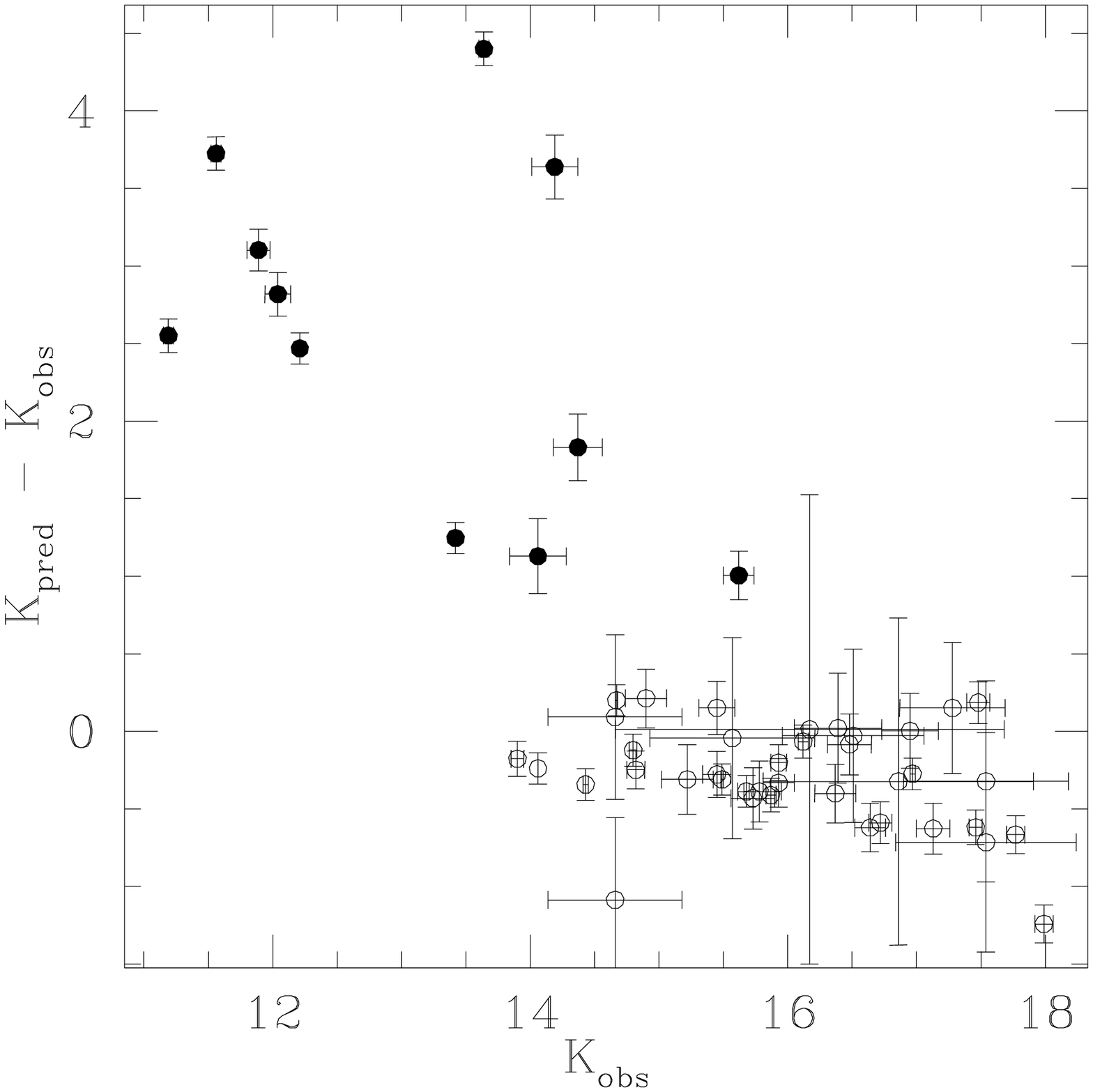}
\caption{Difference between predicted and observed magnitudes
plotted against observed magnitude for IR-observed hot DA white
dwarfs.  Objects for which we determine a significant IR 
excess are shown as filled circles.
\label{f-diff}} 
\end{figure*}

The mean value of $(K_p - K_o)$ for the systems with no IR excess is
$-0.28\pm 0.33$.  We find a similar offset for $(J_p - J_o)$
of $-0.30\pm 0.35$\footnote{The mean of $(J-K)_p - (J-K)_o$ is zero
($-0.03\pm0.29$).}.  We consider several possibilities to explain the
zeropoint offset.  
(1)~{\hbox {\em Our predicted DA (V-K) colors are too red.}} 
Since we derive our DA $K$ magnitudes from the predicted color
$(V-K)_p$ and the published $V$ magnitude, a difference in
normalization in our synthetic $V$ or $K$ photometry of 
the DA models could cause the offset.  However, the synthetic $V$ and $K$
mags and colors we derive for stars in the BPGS spectral library match
their normalizations and colors for each spectral type (see
\S~\ref{kmags} above), and our $(V-K)$
colors also match those in the spectral library of Pickles (1998) to
within a mean difference of $0.01\pm0.1$mag.  
(2)~{\hbox {\em Our IR photometric zeropoint is too faint.}}
Comparing the 2 DAs also observed by Zuckerman \& Becklin (1992), our
$K$ mags are 0.1mag fainter in each case.  While this difference is
well within our errors, we cannot rule out a photometric zeropoint
shift of this order as a contributor to the mean zeropoint
offset.  
(3) {\em Unmodeled, large equivalent width IR absorption
features in DAs.} To account for a shift of 0.3mag in $K$, spectral
absorption features with equivalent widths of nearly 1440\AA\, would
be required that are not present in the DA models.  Given that
different absorption features would be required in the $J$ band to
yield a similar zeropoint shift, we disregard this possibility. 

Even with this unexplained zeropoint shift, our determination of an IR
excess, based on the significance of ($K_p - K_o$), are
unaffected.  However, our estimated companion magnitudes $K_c$ and absolute
magnitudes $M_{K_c}$, may be up to 0.3mag too faint, 
leading in turn to spectral subclass estimates approximately half a
subclass too late.  

The $V$ magnitudes of the candidate companions listed in
Table~\ref{tcompanions} are in most cases significantly fainter than
the visual magnitudes of the DA WDs.  Since we elected not to observe
in the IR those systems known to be composite by their optical
spectra, this result is expected from our selection.   


\subsection{Spatially Resolved Companions}

Finding charts for most of our sample have been published by Shara et
al. (1993, 1997).  To account for visually resolved companions, we
inspected images from the DSS (Shara et al. 1997), and also
searched the USNOA-1.0 (Monet 1996) catalog to $R\sim19.5$ for all
stars within 15\arcsec\, of the program star.  Resolved stars within a
maximum distance of 10\arcsec\, and within 3mag (in $R$) of the
primary are noted in Table~\ref{tsample} as visual pairs.  For the
closest WD distances in our sample of $\sim 20$pc, the spatial
resolution of $\sim 2\arcsec$\, in both the archival optical and new
IR images, corresponds to 40~AU.  On the other hand, a search limited
to 10\arcsec\, corresponds to maximum separations of 200~AU.  For the
farthest WDs in our sample near 500pc, imaging is instead sensitive to
resolved companions between 1000 and 5000~AU.  






\section{Notes on Individual Objects}

REJ\,0134-160: 
Based on radial velocity measurements (Schultz et al. 1996)
and its low mass (NGS99, Finley et al. 1997) mass transfer is unlikely
to have occurred in this binary DA+M star system.

REJ\,0148-253:
Finley et al. (1997) claimed evidence
for a companion based on H$\beta$ emission and a slight red excess.  
Our result of an IR excess confirms the presence of the companion.

\begin{figure*}
\plotone{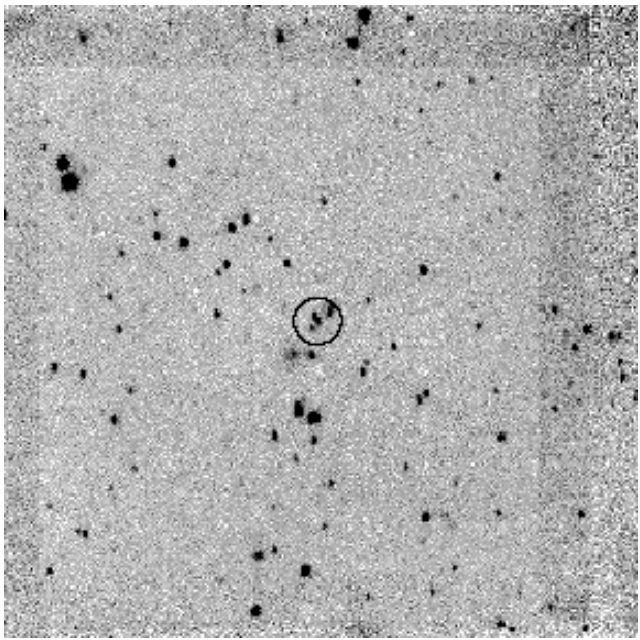}
\caption{A $5\times 5\arcmin$ $K$ band image of the field of REJ\,0427+7406
reveals several close neighbors to the DA.  Diffuse emission 
is also evident to the SE, possible a background galaxy.
North is up, East to the left.
\label{f-re0427p74k}} 
\end{figure*}

REJ\,0427+7406
This star appears to be a multiple at near-IR wavelengths.  Some
nebulosity  (diffuse emission) is also noted nearby.  A finding
chart is provided in Figure~\ref{f-re0427p74k}.

REJ\,0443-034:  
This is the hottest DA in our sample of IR-observed WDs.  DA stars
with such high temperatures should cool extremely rapidly and thus be
detected only rarely.  The suggestion by Finley et al. (1997) of a
possible red excess is not borne out by our measurements.

REJ\,0512-004: 
As the least massive DA in the sample, this DA might be expected to
have a companion, since very low mass WDs are thought to form in
binaries (Marsh, Dhillon, \& Duck 1995).  The absence of IR excess
indicates a very cool and/or degenerate companion at best. 
Our IR photometric method is insensitive to such objects, which
are best detected through radial velocity variations
(Saffer, Livio, \& Yungelson 1998).  We discuss the overall
effect on our estimated binary fraction in
section~\ref{binaryfraction}. 

REJ\,0916-194: While its IR excess suggests a late-type
companion, our estimate for its visual magnitude
$V_c$ is only $\sim2$mag fainter than the total
$V$ mag, so evidence for the companion should have been seen in the red
end of its optical spectrum.  However, the only spectrum (from Vennes
et al. 1997a) is of comparatively low S/N and extends only to 6040\AA.
We expect that a higher S/N optical spectrum extending farther
redward, or our own planned IR spectroscopy will readily reveal
the companion.

REJ\,1036+460 (PG\,1033+464):
$K$ photometry and $K$-excess from Zuckerman \& Becklin (1992) are
within 0.1mag of the values we report here.  However, their derived
absolute red dwarf magnitude $M_{K_c}$ is fainter by 1.5mag.  Rather
than using detailed spectral fits, they had to use published
color-estimated distance moduli that assumed uniform surface gravity
for all white dwarfs, while in fact the absolute magnitude
depends strongly on \logg.  For the other object in common between our
samples, REJ\,1126+183 (PG\,1123+189), we note an even more striking
discrepancy (3.2mag in $M_{K_c}$).  They assume a distance of 30pc,
while the best estimate from detailed fits of NGS is 135pc.  Their
resulting claim that the IMF is flat or increasing down to at least
0.1\msun should be revisited with more accurate distance determinations.

REJ\,1043$+$490:
This was suggested by Schwartz et al. (1995) as a possible multiple
system based on a slightly red $R-I$ color, but they suggested that
further measurements were needed to confirm this.  The residuals of
the DA model fit to the optical spectrum in NGS99 confirmed a red
excess for $\lambda>4500$\AA\ corresponding to a companion with
$V\sim18.2$ or $M_V\sim 10.4$ at our distance of 370pc.  
However, our near-IR photometry is inconsistent with such a companion,
since a normal late M dwarf companion of this absolute magnitude and
distance would yield $J\sim 14.5$ and $K\sim 13.6$.  We note that the
system is a visual binary, with a red companion 8$\arcsec$ West of the
DA. Contamination of the DA spectrum in NGS99 is thus the likely
explanation.

REJ\,1235+233 (PG1232+238):
The object is barely discernible in our $K$ mosaic, so reported
$K$ photometry is poor.

REJ\,1711+664:  This is a barely resolved visual pair,
with a late-type star only $\sim2\arcsec .5$ distant,
which is included in our IR photometry (Mason et al. 1995).  
Jomaron (1997) also found the M-type secondary
in residuals to the DA optical spectrum, classifying the former as type
M3--M4, with no apparent H emission.  Our IR photometry indicates
a somewhat later type.  

REJ\,2207+252: IR photometry for 2207+252 was contaminated by a 
very bright, resolved object at 8.9\arcsec. 

\section{Selection Biases}
\label{bias}

Strong biases are likely to be inherent in any sample of EUV-detected
WDs.  The strongest is the selection against a high column of
intervening interstellar material.  Columns as low as 
$5\times 10^{19}\cmsq$ easily quench flux in the EUV below detection
levels achieved in either the ROSAT WFC or the EUVE.\footnote{For
example, one optical depth is reached at column densities of 30, 5, 1,
and 0.5 $\times 10^{18}\cmsq$ respectively for the EUVE 100, 200, 400,
and 600\AA\, detection bandpasses  (Vennes et al. 1996).}
When compared to more distant areas in the Galactic plane, the local
interstellar medium lacks cool dense material ($n\sim0.1$cm$^{-3}, 
T\sim 100$K).   This `local bubble' has a mean radius of 70~pc
(Diamond et al. 1995). Accurate estimates of the intervening column to
each star are difficult, requiring high S/N EUV spectral analysis
well beyond the scope of this work.  The global effect
amounts to a direction-dependent flux sensitivity limit (and therefore
volume) that is strongly dependent on the fine structure of the ISM.

As pointed out by NGS99, at a given temperature, WDs with lower masses
are larger, consequently more luminous, and thus detectable to larger
distances. For instance, at 30,000\,K, a WD of 0.5\msun\ is detectable
at distances 30\% larger than a 0.7\msun\ WD.  For many WDs at the
maximum distance sampled by most optical surveys, interstellar matter
effectively absorbs all EUV radiation. Thus the dominant effect in EUV
surveys is probably not a selection {\em for} massive WDs as suggested
by V97, but a selection {\em against} low-mass WDs due to a sample
volume strongly affected by interstellar absorption.  However,
there is no direct bias induced by interstellar absorption
on the observed binary fraction.

There have been suggestions that detection of soft X-ray flux from
hot white dwarfs may be biased against close binary systems.
This could arise if heavy elements arising in a wind from
the MS companion have accreted onto the surface of the white dwarf,
enhancing atmospheric EUV and soft X-ray opacity (Barstow et
al. 1993).  By including only DA stars in our sample, we might thus be
preferentially excluding some binary systems.  For instance, the hot
white dwarf in the close binary RE\,1016-053 is a DAO star whose mixed
H/He composition is likely to be attributable to slow accretion from a
red dwarf companion (Vennes et al. 1997b). However, Barstow et
al. (1994) found that hot ROSAT (PSPC or EUV) detected WDs in
non-interacting binaries with early-type stars showed no signs of
abundance enhancements from accretion, their opacity sources following
the same temperature pattern as seen for isolated WDs.  Indeed,
while those abundance enhancements may lead to significant
X-ray opacity (e.g., Dupuis et al. 1995) in the hottest WDs, they
remain strong EUV sources, particularly at $\lambda>200$\AA.

Arguments for biases of opposite effect exist as well - that increased EUV
flux is more likely from a WD that suffers reheating due to accretion
of material from its companion.  Since the DAs in our sample are not
significantly variable, the interaction would have to be fairly weak
or intermittent (e.g. via weak wind accretion).  Furthermore,
Barstow et al. (1992) find for the DA+K2V binary system V471~Tauri
that the temperature change in such cases (with accretion
rates of $0.4-11\times10^{-13}\msun\,{\rm yr}^{-1}$) is small and
occurs only in the region near the hot spot of the accretion column.

Measurement of these putative competing selection effects would
clearly benefit from larger samples. A large number of EUV-detected DA
white dwarfs with good constraints from atmospheric modeling now exist
(Marsh et al. 1997, Vennes et al. 1997a, Finley, Koester \& Basri
1997, NGS99).  The 2MASS will provide (less sensitive, $K\leq14$) IR
photometry for brighter subsamples.  

\begin{figure*}
\plotone{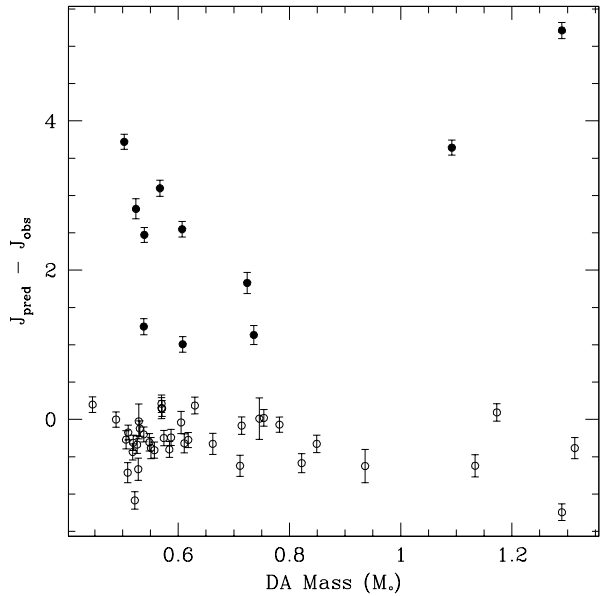}
\caption{$J$ excess (difference between predicted and observed 
$J$ magnitudes) plotted against DA white dwarf mass in solar units
from the best-fit model atmospheres of NGS99. We see no evidence that
IR excess correlates with DA mass.  
\label{f-jdiffmsol}} 
\end{figure*}

WDs with masses less than the canonical core-helium ignition mass
(0.49\msun; Sweigart 1994) are likely to be in short-period binary
systems, where their evolution was cut short by wind ejection during a
common envelope phase (Marsh, Dhillon, \& Duck 1995, Marsh 1995).
Only one of the WD stars in our sample, REJ\,0512-004, has a low mass
(0.45\msun), but it shows no signs of a companion.  On the other hand,
lower mass companions may be preferentially removed by merging, a
process that is more effective the more massive the WD.  

It is thus of interest to check for any correlation between DA mass
and IR excess. Figure~\ref{f-jdiffmsol} shows that no such correlation
is detectable in our sample, as confirmed by two correlation
tests.  Similarly, there is no correlation between the absolute $K$
magnitudes of companions $M_{K_C}$ and those of their corresponding DA
white dwarfs.\footnote{Generalized Kendall`s Tau and Cox
Proportional Hazard model tests from the ASURV package (LaValley, Isobe \&
Feigelson 1992) show probabilities $P$ of 18\% and 72\% respectively
that a correlation is {\em not} present between $(J_p - J_o)$
and DA mass.  A significant correlation requires
$P>$99\%. $P$ of 26\% and 36\% respectively, are obtained for
$M_{K_c}$ vs. $M_{V_{DA}}$.}  Two sample statistical tests also reveal
no significant difference between the mass or absolute
magnitude distributions of the DAs in the no-excess and excess
subsamples.\footnote{Generalized Wilcoxon and logrank tests show that
the two samples are inconsistent with being drawn from the same parent
population at only the $P\sim 75\%$ level for DA mass and
only the $P\sim 93\%$ level for $M_{V_{DA}}$.}   Larger samples would put
better constraints on these hypotheses.

\section{The Binary Fraction}
\label{binaryfraction}

Our WD sample demands that the initial primary has already passed
through the AGB phase, imposing a lower mass limit of about 2\msun\, on
the WD progenitor.  We thus probe portions of the (initial) mass ratio
distribution that are quite difficult to reach in unevolved systems.
The binary frequency of high mass stars has proven difficult to
determine for several reasons.  Photometric detection of a faint low
mass star next to a bright early type primary requires high precision
photometry.  The use of radial velocity variations can also be
difficult due to the large rotational speeds of many early-type stars
- one reason why solar-type stars are preferred in current planetary
searches.

Out of our uniformly-selected sample of 57 EUV-detected DA white
dwarfs, including our new detections of IR excess, we find strong
evidence for multiplicity in 20 systems (35\%) within a 10\arcsec\,
radius.  Simple Poisson statistics on the subsample yields an error of
7\%.  We may also count multiple systems by assuming that any resolved
object within the 10\arcsec\, radius is at the same distance as the WD
(i.e., a presumed companion), and thereby tally objects within a
chosen limiting physical separation (projected onto the sky
plane). Since the largest estimated distance to any of
the WDs in our sample is 470pc (for REJ\,1235+233) we choose for completeness
\footnote{Assumes a spatial resolution of 2\arcsec\, at a 
similar maximum distance of 500pc.  Including {\em only} unresolved
systems is intuitively appealing, but introduces a bias against
companions in nearby systems.} a maximum projected physical
separation of 1000AU. This removes only one system (1726+583), leaving
the binary fraction essentially unchanged at $33\pm7$\%. 

For comparison, among a complete sample of nearby ($d<5$pc) M dwarfs,
Leinert (1997) found $26\pm 9\%$ to have companions.  In a
meta-analysis, Fischer \& Marcy (1992) found that $42\pm9 \%$ of M
dwarfs have main sequence companions.  This is lower than the best
incompleteness-corrected estimate for G dwarfs ($\sim 57\%$; Duquennoy
and Mayor 1991), due in part to the smaller range of secondary
companion masses available.

As allowed by the faintness of WDs relative to main sequence
companions, and the use of IR photometry, the current study is
sensitive to a wide range of companion masses, from late M to early
A-type.  The late-type companion masses corresponding to our estimates
of absolute $K$ magnitudes $M_{K_C}$ range from about 0.5 down to
$0.1\msun$.\footnote{Below about 0.2\msun, absolute magnitudes no
longer correlate well with mass.  There the relationship steepens
considerably, in both $V$ and $K$ bands, and differs little as a
function of metallicity (Baraffe et al. 1998; Henry et al. 1999).}
Judging by known or estimated spectral types of unresolved companions
and by derived absolute magnitudes of the presumed companions
discussed above, 90\% of companions from Tables~\ref{tsample} and
\ref{tcompanions} are of spectral type K or later.  

We note that despite the significant improvement in sensitivity to low
mass companions that our IR photometry affords, the binary frequency of
our sample remains a lower limit for several reasons.  First, very low
mass companions can fall below our IR detection limit of $K\sim16$,
with sensitivity decreasing with distance.  Also, while we assume only
a single degenerate in each system, double degenerates would
not likely be distinguished by our IR photometry.  Saffer, Livio, \&
Yungelson (1998) derive a fraction of double degenerate systems 
that may be as high as 20\% when they include an estimated 
correction factor of $\sim1.4$ for selection efficiency in their
sample.   

Some of the hot DAs with masses higher than expected from isolated evolution
($\sim 1.1\msun$) have been hypothesized to represent a population of
coalesced double degenerate systems (e.g., Marsh et al. 1997).  Such
mergers would slightly reduce the observed binary fraction in our
sample.  While they might also help constrain the Type~Ia supernova
formation rate (e.g., Branch et al. 1995), the predicted peak of such
a merger distribution  should be near $0.9\msun$, which is not seen. 

Among our sample of hot DAs may be some close binaries that have
passed through a phase of common envelope evolution.  While these are
sometimes called precataclysmic binaries, many will not become
cataclysmic variables within a Hubble time (Ritter 1986).  Based on
morphological classification of planetary nebulae, and the assumption
that axisymmetrical mass loss is caused by binary companions, Soker
(1997) estimated that only 10\% of PN had no interaction with any
companion, stellar or substellar.  Only 11\% had interactions while
avoiding a common envelope phase.  For the case of close binaries, the
observed distribution of companion masses may thus have been modified 
by evolution.  Some systems may have coalesced, while others may have
grown through wind accretion, or by stable mass transfer.  As the
progenitors of novae, CVs, symbiotics, Feige~24-type systems, and the
intriguing dwarf carbon (dC) stars (deKool \& Green 1995), these WD+MS
systems warrant the improved constraints on their masses and abundance
enhancements that IR spectroscopy of our sample will provide.  For
instance, the dC stars, which likely evolve into CH or Ba giants, are
dwarfs whose strong carbon bands betray episodes of mass-transfer from
the previously-AGB companion, now the WD (Green 2000).  There is
recent evidence (Heber et al. 1993, Liebert et al. 1994) that dCs may
show stronger abundance enhancements soon after the AGB phase, i.e.,
while the WD is still hot. IR spectra would test for this, and its
potential ramifications for the depths and timescales of mixing due
to chemical potential gradients (e.g., Proffit \& Michaud 1989).

We are obtaining IR spectra of these systems at the KPNO 2.1m
with CRSP to better constrain the mass and spectral type of the cool
companions presumably causing the IR excess.   We await these
results before pursuing more detailed discussions of the mass function
of the companions, the observed final and implied
initial mass ratios of the systems, and detailed simulations
of the effects of binary evolution.   The binary fraction may be a
function of the mass ratio, or of the primary mass; these fundamental
questions of stellar evolution remain open.  If the observed binary
fraction for DA WDs proves to be inconsistent with those of main
sequence primaries in a similar (initial) mass range, it could mean
that age-dependent formation conditions (e.g., abundance, kinematics,
or turbulence) strongly influence the binary fraction.

Many thanks to Dick Joyce and the staff at KPNO for all their
help during IRIM observing runs. Thanks to Fred Ringwald for
providing a catalog of unresolved, hot-high-gravity with cool
companion composites. The author gratefully acknowledges
support provided by NASA through Grant NAG5-1253, and Contract
NAS8-39073 (CXC).  This research has made extensive use of SIMBAD.

\clearpage
\appendix 

\section{Photometry of mosaiced images} 
\label{apx-phot}

\begin{figure*}
\plotone{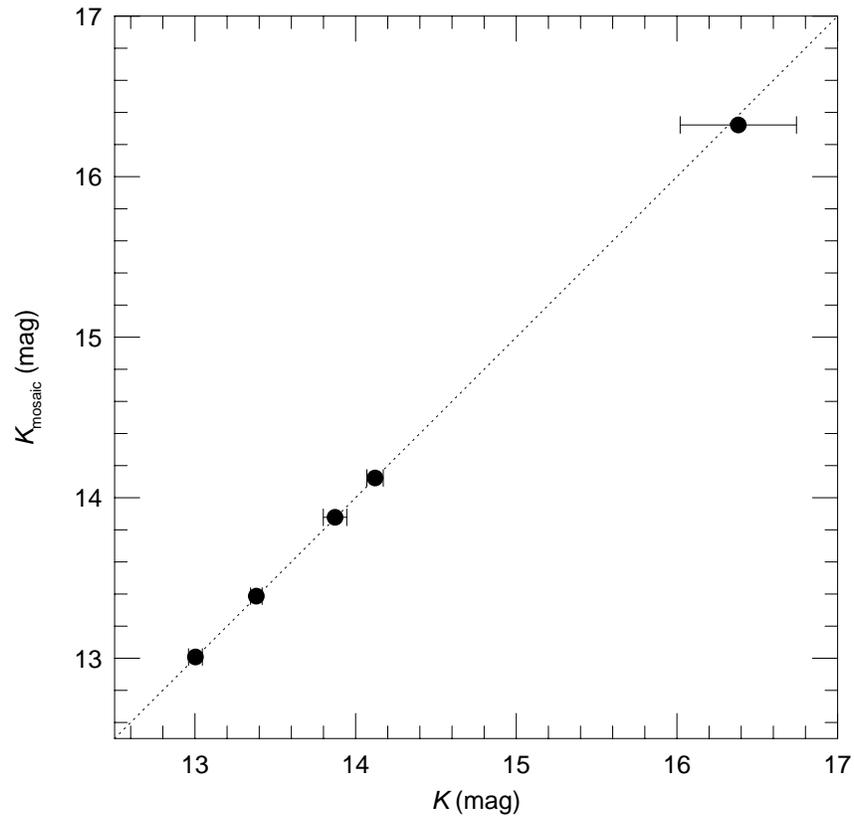}
\caption{A comparison of the derived magnitudes in the field of
PG~0904+511 using individual frames and a mosaic image.
{\it K}-band magnitudes derived from an image mosaic and 
those from individual frames are compared for 5 stars in the
PG~0904+511 field.   The error bars show the 1-$\sigma$\ deviation
among magnitudes from individual frames.   The dashed line
shows the resulting relationship if the two quantities were
identical.
\label{f-mosphot}} 
\end{figure*}

For some targets, visual inspection failed to reveal the source on the
individual images.  We derive photometry these targets from
a global mosaic of the field produced by shifting and combining
individual frames.  We use integer pixel shifts) and average the
overlap regions between frames, accounting for background/sky
variations by subtracting the median value of the frame from itself
before averaging the fields.  Integer pixel shifts do not necessarily
conserve flux, but are employed here for two
reasons: (1) the stellar images on our frames are sub-Nyquist sampled.
And (2) the presence of significant dead (insensitive) space between
pixels.  The undersampled profiles of the star generally ensure that
most of the stellar flux is contained within one pixel, and the
presence of dead space combined with undersampled images makes it
difficult to produce flux conserved stellar profiles in the first
place.  Thus we conclude that mosaicing images will not significantly
alter our photometry.  These assumptions are tested by comparing
magnitudes derived from a mosaiced image to those from
individual frames.  Figures~\ref{f-mosphot} illustrated this
comparison.  Figure~\ref{f-mosphot} demonstrates that photometry
from mosaiced images is consistent (within the errors)
with photometry obtained from individual frames.

\begin{figure*}
\plotone{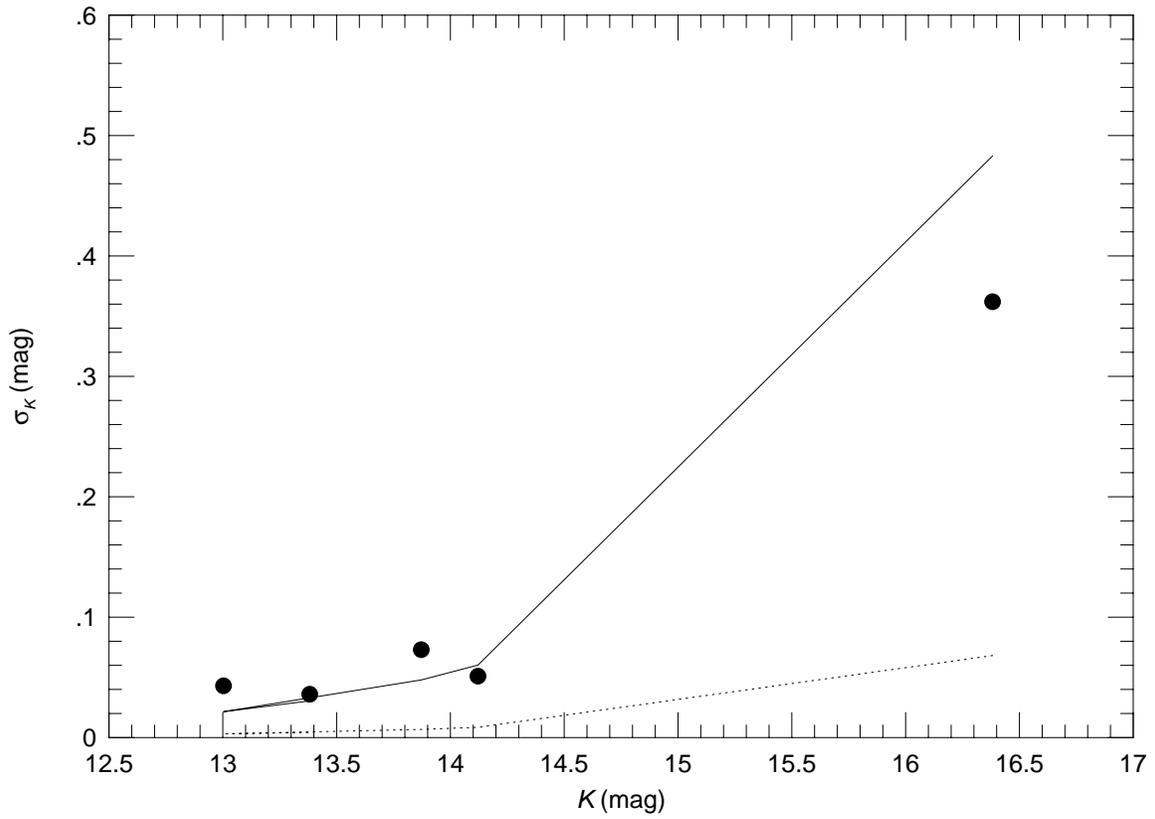}
\caption{The observed error (estimated as the 1-$\sigma$\ deviation
among magnitudes derived from individual frames) {\it versus}\
magnitude for the PG 0904+511 field.  The dashed line shows the
magnitude of the formal errors from S/N arguments.  The
solid line shows our error curve for the field (see text for
details).  These data suggest that we are dominated by systematic
errors.
\label{f-moserr}}
\end{figure*}

The mosaiced images, however, only allow us to make an estimate of the
formal (Poisson + detector noise) errors, and do not account for any
systematic errors or any additional possible degradation 
of the S/N created by the photometry software.
A comparison of the formal errors with those obtained by comparing
the RMS magnitude variation from individual frame suggests 
that our errors are dominated by these systematic errors.
The dashed line in
Figure~\ref{f-moserr} shows the magnitude of the formal errors.  The
solid dots are the observed RMS variations of magnitudes from
individual frames.  Clearly, formal errors underestimate the measured
empirical estimates.  The formal errors are derived using a
modified form of the IRAF task ccdtime (modified to run on the IDL
package).  To obtain a realistic estimate for the errors on mosaiced
images, we modeled the effects of systematic errors by artificially
increasing the readout noise of the detector and minimizing the
relationship:

$$ \sum_{i=1}^{N}\frac{(\sigma_{obs}-\sigma_{cal})^2}{\sigma_{obs}^2} $$

where
$\sigma_{obs}$ are the observed errors, 
$\sigma_{cal}$ are errors calculated by increasing readout noise,
and 
N is the number of stars in this experiment.
The solid line in Figure~\ref{f-moserr} shows an example of this
fit.  We use the best-fit ``readout noise'' to extrapolate the formal
S/N to stars for which only mosaiced image
photometry is available.

\end{document}